\documentclass[aps,pra,twocolumn,showpacs,groupedaddress,superscriptaddress]{revtex4-1}  % for review and submission
\usepackage{graphicx}  % needed for figures
\usepackage{dcolumn}   % needed for some tables
\usepackage{bm}        % for math
\usepackage{amssymb}   % for math

\usepackage{float}
\usepackage{verbatim}   % useful for program listings

\usepackage{amssymb}
\usepackage{amsmath}
\usepackage{amsfonts}
\usepackage{color}
\usepackage{framed}
\usepackage{dcolumn}
\usepackage{ulem}
\usepackage{framed}
\usepackage{verbatim,moreverb,bm}
\usepackage{graphicx}
\usepackage{amsmath}
\usepackage{float}
\usepackage{esvect}
\usepackage{braket}
\usepackage[autostyle]{csquotes}

\DeclareMathOperator{\Tr}{Tr}
\let\Re\undefined
\DeclareMathOperator{\Re}{Re}
\let\Im\undefined
\DeclareMathOperator{\Im}{Im}

\usepackage{soul}
\usepackage{epstopdf}

\begin{document}

\widetext
%\leftline{Version 10 as of \today}
%\leftline{Primary authors: }
%\leftline{To be submitted to PRA}
%\leftline{Comment to {\tt d0-run2eb-nnn@fnal.gov} by xxx, yyy}
%\centerline{\em D\O\ INTERNAL DOCUMENT -- NOT FOR PUBLIC DISTRIBUTION}

% the following line is for submission, including submission to the arXiv!!
%\hspace{5.2in} \mbox{Fermilab-Pub-04/xxx-E}

\title{Sub-Doppler Laser Cooling using Electromagnetically Induced
  Transparency}

\author{Peiru He}
\affiliation{%
JILA and Department of Physics, University of Colorado, Boulder, Colorado 80309-0440, USA
}%
\affiliation{%
Center for Theory of Quantum Matter, University of Colorado, Boulder, Colorado 80309, USA
}%
\author{Phoebe M. Tengdin}
\affiliation{%
JILA and Department of Physics, University of Colorado, Boulder, Colorado 80309-0440, USA
}%
\author{Dana Z. Anderson}
\affiliation{%
JILA and Department of Physics, University of Colorado, Boulder, Colorado 80309-0440, USA
}%
\author{ Ana Maria Rey}
\affiliation{%
JILA and Department of Physics, University of Colorado, Boulder, Colorado 80309-0440, USA
}%
\affiliation{%
Center for Theory of Quantum Matter, University of Colorado, Boulder, Colorado 80309, USA
}%
\author{Murray Holland}
\affiliation{%
JILA and Department of Physics, University of Colorado, Boulder, Colorado 80309-0440, USA
}%
\affiliation{%
Center for Theory of Quantum Matter, University of Colorado, Boulder, Colorado 80309, USA
}%

%\input author_list.tex       % D0 authors (remove the first 3 lines
                             % of this file prior to submission, they
                             % contain a time stamp for the authorlist)
                             % (includes institutions and visitors)
\begin{abstract}
  We propose a sub-Doppler laser cooling mechanism that takes
  advantage of the unique spectral features and extreme dispersion
  generated by the phenomenon of electromagnetically induced
  transparency (EIT). EIT is a destructive quantum interference
  phenomenon experienced by atoms with multiple internal quantum
  states when illuminated by laser fields with appropriate
  frequencies.  By detuning the lasers slightly from the ``dark
  resonance'', we observe that, within the transparency window, atoms
  can be subject to a strong viscous force, while being only slightly
  heated by the diffusion caused by spontaneous photon scattering.  In
  contrast to other laser cooling schemes, such as polarization
  gradient cooling or EIT-sideband cooling, no external magnetic field
  or strong external confining potential is required.  Using a
  semiclassical approximation, we derive analytically quantitative
  expressions for the steady-state temperature, which is confirmed by
  full quantum mechanical numerical simulations.  We find that the
  lowest achievable temperatures approach the single-photon recoil
  energy. In addition to dissipative forces, the atoms are subject to
  a stationary conservative potential, leading to the possibility of
  spatial confinement. We find that under typical experimental
  parameters this effect is weak and stable trapping is not possible.

\end{abstract}

\maketitle

\section{Introduction}

Techniques for laser cooling and
trapping of atoms have facilitated major advances in quantum
science, and are now opening a window to the use of atomic systems for
quantum information processing
tasks~\cite{RevModPhys.70.707,RevModPhys.70.685,RevModPhys.70.721}.
Examples of these developments include the recent demonstrations of
Bose-Einstein condensation and quantum degenerate Fermi gases, the
coherent manipulation of individual atoms for the implementation of
quantum logic~\cite{PhysRevLett.75.4714}, and the quantum simulation
of model Hamiltonians such as the Bose-Hubbard
model~\cite{ISI:000173028800033}. 

Despite the evident success of laser cooling methods,
a variety of complex phenomena predicted in studies
of strongly-correlated materials
call for lower atomic temperatures than standard laser cooling techniques provide.. 
For this reason, there is substantial interest in new laser cooling
techniques that can achieve lower temperatures, higher densities, more rapid cooling, 
and be applicable to more general systems.

Doppler cooling is perhaps the most basic and fundamental of the laser
cooling methods used in quantum gas experiments.  This technique can
cool two-level system to a temperature of the order
of $\hbar\Gamma/k_B$~\cite{Castin:89}, where $\Gamma$ is the natural
linewidth of the atomic transition, and $\hbar$ and $k_B$ are the
reduced Planck's constant and Boltzmann's constant, respectively.  The
validity of this expression is limited to the case when~$\hbar\Gamma$
exceeds the recoil energy, given by $E_r=\hbar^2k^2/2m$, with
$m$ the atom's mass and $k$ the transition's wavevector. In order to
reach sub-Doppler temperatures, other laser cooling methods such as
Sisyphus cooling~\cite{Dalibard89, Wineland:92} can be used that take
advantage of a multilevel internal atomic structure. These typically
allow the energy of an atom to be cooled to be of the order of the
recoil energy~\cite{RevModPhys.70.707}.  Similar sub-Doppler cooling temperature
have been achieved using a three-level configuration by
using two pairs of standing waves
~\cite{PhysRevLett.71.3087, fernandes2012sub, grier2013lambda, sievers2015simultaneous, morigi2007two}.

Laser cooling techniques that achieve sub-recoil temperatures have
generally required the use of dark-state resonances.  These include
velocity selective coherent population trapping
(VSCPT)~\cite{PhysRevLett.61.826}, as well as recently proposed
techniques that combine electromagnetically induced transparency
(EIT)~\cite{PhysRevLett.66.2593} with sideband
cooling~\cite{PhysRevLett.85.4458}.  The latter have enabled
experimentalists to efficiently cool fermions in quantum gas
microscopes~\cite{haller_single-atom_2015,PhysRevA.92.063406}.

\begin{figure}[H]
\centering
\includegraphics[scale=0.5]{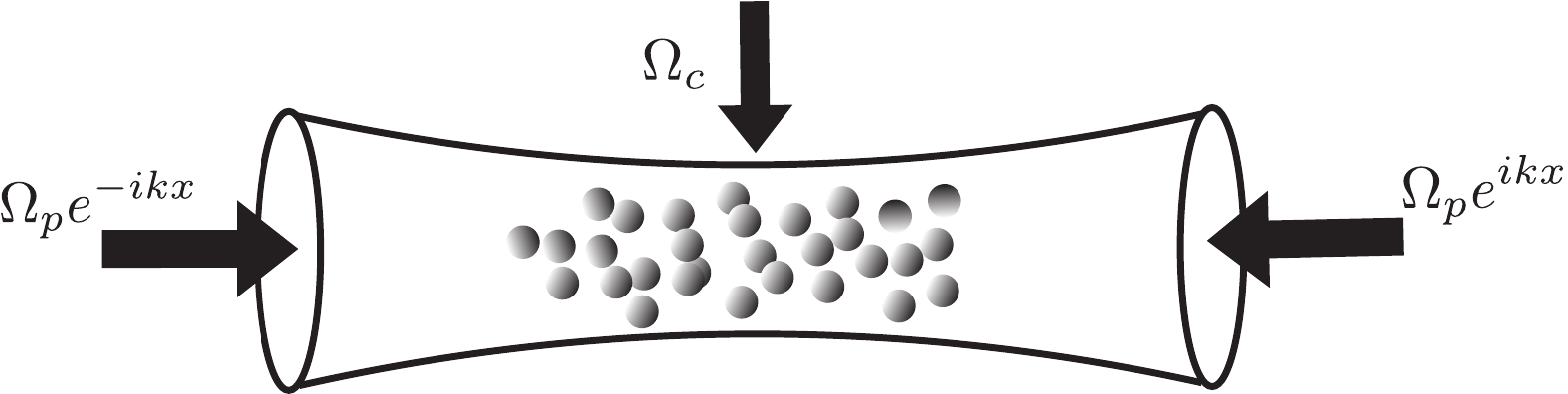}
\caption{The system we consider consists of atoms transversely
  confined so that they are able to move only in the $x$ direction.
  The probe lasers consist of a pair of counter-propagating beams
  aligned with the $x$-axis with equal intensity as characterized by
  the Rabi frequency $\Omega_p$. The coupling laser propagates
  perpendicularly to the probe beams as shown and has intensity
  characterized by the Rabi frequency $\Omega_c$.  }
\label{fig:setup}
\end{figure}
In this paper, we propose and analyze a laser cooling technique that
combines the benefits of Doppler-cooling in multi-level atoms (in our
case a three-level lambda system) with EIT.  
The distinctive dispersion relations of the EIT system have been well exploited in non-linear optics,
such as in the demonstration of extreme slow light in ultracold atomic gases~\cite{hau1999light}.
The technique we use builds upon the well-known features of EIT to provide 
laser cooling to the recoil energy limit.
Moreover, working with induced transparency suggests that
this technique may provide effective cooling of high-density atomic clouds, 
a regime that challenges standard laser cooling methods
~\cite{gallagher1989exoergic, julienne1989collisions, 
walker1990collective}.
As depicted
in Fig.~\ref{fig:setup}, atoms are illuminated by two
counter-propagating probe beams and an additional coupling beam
directed perpendicularly to the other two. We show that in this
system, laser cooling is possible when the probe lasers are slightly
blue-detuned from the ``dark resonance''.  As in all laser cooling
methods, the equilibrium temperature is determined by a balance
between the heating rate induced by spontaneous emission and the
cooling rate generated by dissipative forces.  The existence of a
dark-state, together with the unique dispersion relations of the EIT system,
modifies both rates: significantly reducing the heating rate by
suppressing absorption, and simultaneously weakening dissipative
forces. On balance, we find that the cooling achieved by the scheme
can allow the atoms to reach final temperatures
approaching the recoil energy limit.  

In addition to these dissipative and fluctuating forces,
we find the atom-probe interaction also imposes a periodic
conservative optical potential. We find that the modulation depth of this
conservative potential is comparable to the achievable temperatures,
leading to the possibility of weak transient confinement and trapping
that can significantly modify the dynamics.

The outline of the paper is the following: In Sec.~\ref{sec:setup} we
introduce the proposed scheme including the atomic level structure and
the laser configuration. We then present a brief review of the
essential aspects of the EIT phenomenon and how this can be exploited
to provide sub-Doppler cooling.  In Sec.~\ref{sec:mechanism}, we
discuss the physics behind the cooling mechanism and determine the
achievable final temperatures.  We use two different approaches: a
semiclassical treatment that allows us to analytically derive
expressions for the cooling rate, the capture range, the diffusion,
and the equilibrium temperature; and a quantum Monte-Carlo wave
function (MCWF) treatment which we use to numerically determine the
capability of the proposed scheme to reach recoil-limited final
temperatures.  In Sec.~\ref{sec:trapping}, we consider the atomic
motion and analyze the resulting spatial diffusion.
Sec.~\ref{sec:conclusion} provides concluding remarks..

\section{The Setup}
\label{sec:setup}

\begin{figure}[H]
\centering
\includegraphics[scale=0.5]{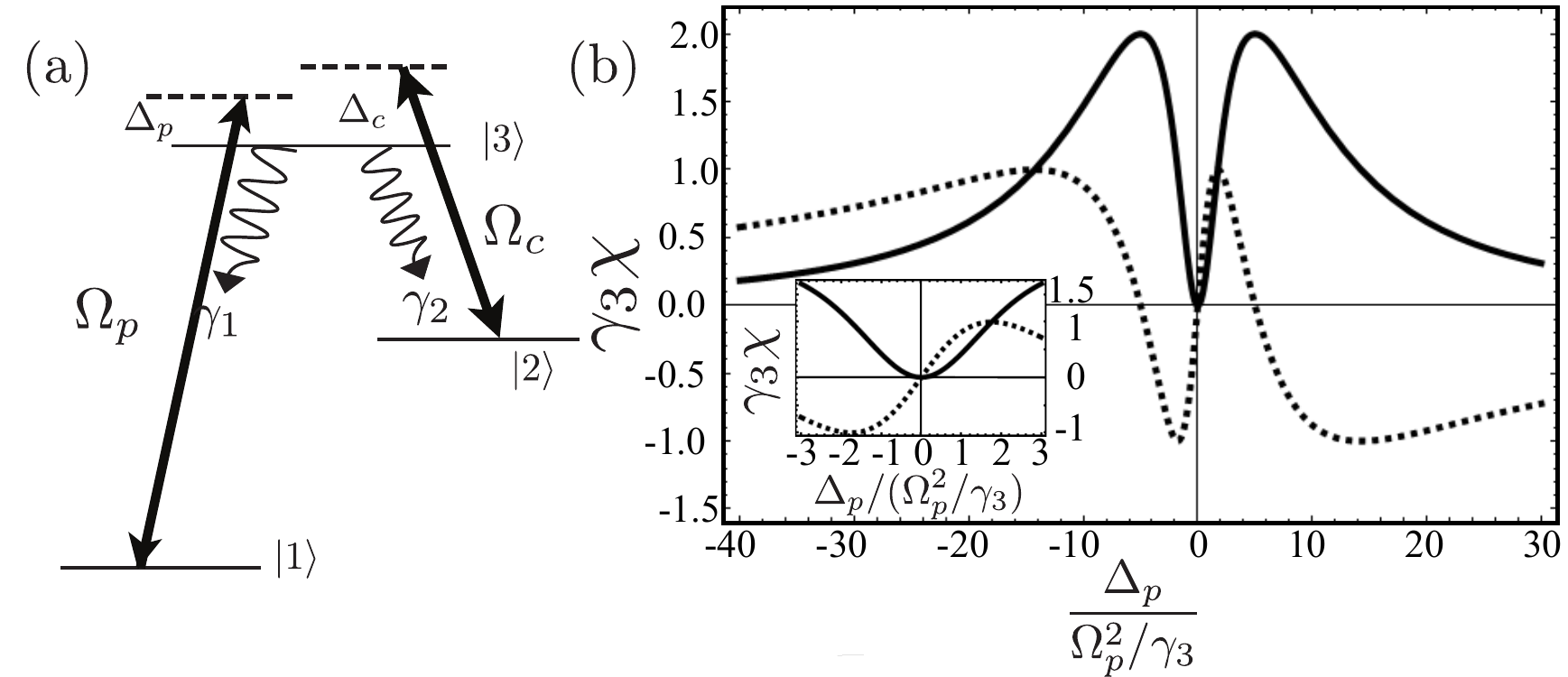}
\caption{(a) Lambda-type three-level atoms driven by two laser beams.
  (b) The real (solid line) and imaginary (dashed line) parts of the
  susceptibility $\chi$ are shown as a function of detuning $\Delta_p$
  for the case $\Delta_c=0$. The inset magnifies the small detuning
  region to highlight the behavior of the susceptibility in the
  transparency window.  
  The parameters used for the plot are, $\Omega_p=10E_r/\hbar$,
  $\Omega_c=400E_r/\hbar$ and $\gamma_3=2000E_r/\hbar$.}
\label{fig:figure1}
\end{figure}

Consider a stationary atom with three internal states, as shown in
Fig.~\ref{fig:figure1}(a).  The excited state, $|3\rangle$, can decay
to two stable ground states, $|1\rangle$ and $|2\rangle$, with decay
rates $\gamma_1$ and $\gamma_2$, respectively.  The total decay rate
of the excited state is thus $\gamma_3 \equiv \gamma_1+\gamma_2$. The dipole
allowed transition $|2\rangle\rightarrow |3\rangle$ with transition
frequency $\omega_{23}$ is driven by a coupling laser with frequency
$\omega_c$, detuned from the atomic transition by
$\Delta_c= \omega_c-\omega_{23}$, and with intensity characterized by
the Rabi frequency $\Omega_c$.  Similarly, the transition
$|1\rangle\rightarrow |3\rangle$, with transition frequency
$\omega_{13}$, is driven by a probe laser with frequency $\omega_p$,
detuned from the transition frequency by
$\Delta_p= \omega_p-\omega_{13}$ and with Rabi frequency $\Omega_p$.

This three-level scheme is a standard $\Lambda$-type EIT system.  In
our discussion, we denote the quantum projection operator
$\hat \sigma_{ij}=|i\rangle\langle j|$, and the density matrix
elements $\rho_{ji}=\langle\hat \sigma_{ij}\rangle$. The
susceptibility for the probe laser field, defined as
$\chi=\rho_{13}/\Omega_p$, is a function of the detuning~$\Delta_p$.
Its real part, $\text{Re}[\chi]$, characterizes refraction of the
probe light, while its imaginary part, $\text{Im}[\chi]$,
characterizes absorption.  When the two-photon resonance condition,
$\Delta_p-\Delta_c=0$, is fulfilled, there exists a perfect
dark-state, $|\text{Dark}\rangle=\Omega_p|2\rangle-\Omega_c|1\rangle$,
for which the $\text{Im}[\chi]$ vanishes and there is no absorption at
all.  In Fig.~\ref{fig:figure1}(b), we plot $\chi(\Delta_p)$ (solid
and dashed lines) in the parameter region
$\Omega_p, \Omega_c\ll\gamma_3$ with the assumption that
$\Delta_c=0$.  Centered around the zero absorption point,
$\Delta_p=0$, is a detuning window of characteristic width
$2(\Omega_p^2+\Omega_c^2)/\gamma_3$ where absorption is significantly
suppressed.

Although the susceptibility properties are well-understood for
three-level EIT systems, they have not to our knowledge been fully
applied to the case of the laser-cooling of free-space atoms.  A
feature of particular interest in the context of laser cooling is the
potentially reduced heating rate due to suppressed spontaneous
emission in the absorption dip, as compared to the one seen in the
absence of EIT.  In this work, we study the cooling process in the
simplest case of a one-dimensional system consisting of an ensemble
whose atoms can only move along the $x$-axis.
Let us now consider the effects of atom motion and the effects of the spatial
variation of the applied fields.

We assume that the probe consists of two $z$-linear-polarized
counter-propagating laser beams with equal frequency $\omega_p$
traveling in the positive and negative $x$-directions. The coupling
beam is a $z$-polarized traveling wave propagating along the $y$
direction with frequency $\omega_c$.  The total electric field in the
$z$ direction is therefore given by
$E_z=2E_{p} e^{i\omega_{p}t} \cos{kx}+E_c e^{i\omega_{c}t}+{\rm c.c.}$,
where $k=\omega_{p}/c\approx \omega_{13}/c$ is the
wavevector of the probe. In the interaction picture, the effective probe Rabi frequency
$2\Omega_{p}\cos{(kx)}$ is therefore position dependent, while the
coupling beam Rabi frequency~$\Omega_c$ is constant in space.  Both
$\Omega_p$ and $\Omega_c$ are proportional to the corresponding
electric field amplitudes $E_p$ and $E_c$, which, without loss of
generality, can be taken to be real.

For the sake of simplicity we consider the case of a sufficiently
dilute gas so that interactions (collisions, photon reabsorption,
radiation pressure, etc.) can be neglected. Consequently the dynamics
can be modeled using a single-atom Hamiltonian.  The Hamiltonian
$\hat H$ consists of two terms, the external part $\hat H_{\text{ext}}$ that
describes the external degrees of freedom and accounts for atomic
motion, and the internal part $\hat H_{\text{int}}$ that describes the
internal atomic levels.  We will focus on the special case
$\Delta_c=0$ and $\gamma_2=0$, although it should be emphasized that
cases with $\Delta_c\neq 0$ and $\gamma_2\neq 0$ yield qualitatively
similar results with respect to the final temperature.  In the
presence of dissipation, the evolution of the system is described by
the Born-Markov quantum master equation written in terms of the
density matrix $\hat\rho$:
\begin{eqnarray}
\hat H&=&\hat H_{\text{ext}}+\hat H_{\text{int}}\nonumber\\
\hat H_{\text{ext}}&=&\frac{\hat p^2}{2m}\nonumber\\
\hat H_{\text{int}}&=&\hbar\Delta_p\hat\sigma_{11}+
\hbar\left[2\Omega_{p}\cos{(k\hat x)}\hat\sigma_{13}
+\Omega_c\hat \sigma_{23}+\mbox{h.c.}\right]\nonumber\\
\frac{d\hat\rho}{dt}
&=&\frac{i}{\hbar}[\hat\rho, \hat H]+\gamma_3
\int_{-1}^{1}du\,\mathcal{N}(u)\mathcal{L}[
\hat \sigma_{13}e^{iku\hat x}]\hat\rho
\label{eq:EIT}.
\end{eqnarray}
where the Lindblad superoperator
$\mathcal{L}[\hat O]\hat\rho=\frac{1}{2}(2\hat O\hat\rho\hat
O^\dagger-\hat O^\dagger\hat O\hat\rho-\hat\rho\hat O^\dagger\hat O)$
accounts for dissipative processes.  Here, $\mathcal{N}(u)$
parametrizes the normalized dipole radiation pattern projected along
the $x$ axis~\cite{PhysRevLett.116.153002}.

\section{Cooling Mechanism}
\label{sec:mechanism}

\subsection{Semiclassical treatment for a weak probe field}

We break the analysis into separate considerations of two parameter
regimes; the weak probe, $\Omega_p\ll \Omega_c\ll\gamma_3$, and the
strong probe, $ \Omega_c\ll\Omega_p\ll\gamma_3$. Here we begin with
the weak probe regime where the cooling mechanism can be more easily
understood and where it is possible to derive analytic expressions for
the equilibrium temperature and the capture range ({\it i.e.}\ the
range of atomic velocities that can be efficiently cooled).  The
discussion of the strong probe beam case follows in the next section
along with an analysis of the resulting impact of the optical fields
on the dynamics of the atomic motion.

In the weak probe regime, the cooling mechanism is similar to Doppler
cooling: a moving atom asymmetrically absorbs photons from the two
probe beams due to the influence of the Doppler effect.  If the atom
is moving to the right with velocity $v$, it sees a Doppler shifted
frequency from the left propagating beam $\omega_p+k v$, giving an
effective detuning $\Delta_p+k v$, and from the right propagating
beam, a frequency $\omega_p-k v$ and associated detuning
$\Delta_p-k v$.  In an absorption event, the atom's momentum is
shifted by the momentum of the photon, $+\hbar k$ for the right
propagating beam and $-\hbar k$ for the left propagating beam.  On the
other hand, when the excited atom undergoes spontaneous emission, the
direction of the emitted photon is drawn randomly from a dipole
radiation pattern, and the atom recoils in the opposite direction to
the emitted photon. It is the net momentum transfer of absorption and
emission cycles that cools or heats the atom.

In order to cool, it is necessary for the atom to preferentially
absorb photons from the beam that is opposing its motion, rather than
the co-propagating beam. In the ordinary Doppler cooling of a two-level
atom, the absorption rate increases as the detuning approaches zero
and is maximum on resonance. Thus, in order for Doppler cooling to
function, it is required to detune to the {\it red} (lower frequency)
of the resonance, so that the effect of the Doppler shift is in the
correct direction. However, in our case (illustrated in
Fig.~\ref{fig:setup}), the transparency window has an inverted
dependence on detuning and the absorption completely vanishes on
resonance. Thus, by analogy, in this system one would expect that it
is necessary for the probe beams to be detuned to the {\it blue} of
the resonance for laser cooling to occur.
 
In order to describe the motional dynamics while accounting for EIT
interference effects, we develop a semiclassical approach that treats
the internal levels fully quantum mechanically, while expressing the
atom's position and momentum by a classical phase-space coordinate,
$(x,p)$, derived from the averages $p=\langle\hat p\rangle$ and
$x=\langle\hat x\rangle$. The semiclassical approximation requires the
temperature to be sufficiently large that the typical atomic kinetic energy
greatly exceeds the recoil energy.

The force seen by the atom depends on the susceptibility. The simplest
expression occurs for the case of an atom at rest, which reaches a
steady-state susceptibility given by
\begin{eqnarray}
  \chi(\Delta_p)\approx
  \frac{\Delta _p}
  {(\Omega _c^2-\Delta_p^2)-i \Delta _p\gamma_3/2}
\end{eqnarray}
where the approximation $\Omega_p\ll\Omega_c$ has been used
corresponding to the weak probe limit. If the atom is not at rest, but
is moving with velocity $v=p/m$, the atom is effectively illuminated
by two plane waves with different detunings $\Delta_p\pm kv$, split by
the Doppler shift.  In the weak probe regime, $\rho_{13}$ can be
derived perturbatively to lowest order in $\Omega_p/\Omega_c$ by
superimposing these two counter-propagating beams separately
\begin{eqnarray}
\rho_{13}\approx \Omega_p[e^{-ikx}\chi(\Delta_p-kv)+e^{ikx}\chi(\Delta_p+kv)]
\label{eq:coherence}
\end{eqnarray}
The force on the atom is defined as the negative spatial derivative of
the Hamiltonian. 
\begin{equation}
\frac{dp}{dt}=\frac{ d\langle\hat p\rangle}{dt}
=-\langle \frac{\partial\hat H}{\partial \hat x}\rangle
=4\hbar k\sin(kx)\Omega_p\Re[\rho_{13}]
\end{equation}
Substituting the expression for $\rho_{13}$ as given in
Eqn.~(\ref{eq:coherence}) yields
\begin{eqnarray}
\frac{dp}{dt}&\approx&4\hbar k|\Omega_p|^2\sin(2kx)
\Re[\chi(\Delta_p)]\nonumber\\
&&{}-4\hbar|\Omega_p|^2 kv(1+\cos(2kx))\partial_{\Delta_p}\Im[\chi(\Delta_p)]
\label{eq:semi}
\end{eqnarray}
where we have assumed the ultracold limit $kv\ll\Delta_p$.

The first term in Eqn.~(\ref{eq:semi}) is the conservative dipole
force, which averages to zero over a wavelength and which we neglect
because its magnitude is small in the weak probe limit under
consideration.  This force, however, is relevant for strong probe
fields and will be discussed in Sec.~\ref{sec:trapping}.
\begin{figure}[H]
\centering
\includegraphics[scale=0.35]{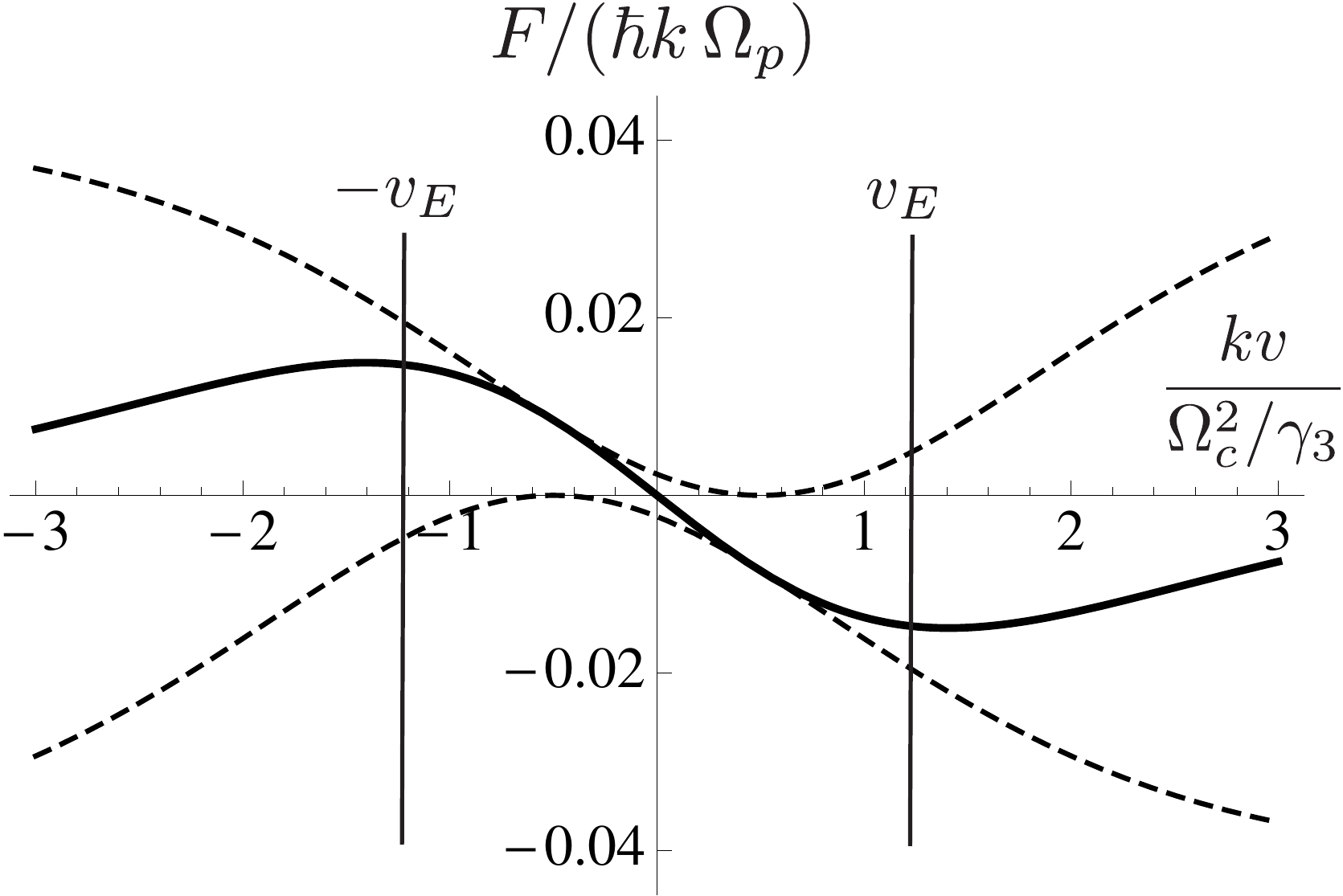}
\caption{Friction forces as a function of atom velocity.  The two
  radiation forces (dashed lines) exerted by the two
  counter-propagating beams give rise to the net friction force (solid
  line).  For blue-detuned probe beams, atoms with positive velocities
  experience a negative force and vice versa, corresponding to
  cooling.  The magnitude of the friction force is maximized for atoms
  traveling at the critical velocity $\pm v_E$ shown, which determines
  the capture range. The parameters used were $kx=0$, $\Delta_p=40E_r/\hbar$, 
  $\Omega_p=20E_r/\hbar$,
  $\Omega_c=400E_r/\hbar$ and $\gamma_3=2000E_r/\hbar$.}
\label{fig:force}
\end{figure}
The second term in Eqn.~(\ref{eq:semi}) is the radiation force which
gives rise to a velocity-dependent dissipative force, as shown in
Fig.~\ref{fig:force}. For small absolute velocities, the radiation
force is approximately linear, and the negative of the slope gives
rise to the friction coefficient, denoted by~$\eta(x)$.  As the
velocity increases, the dependence of the force on velocity becomes
non-linear and a complicated expression is needed to describe the
friction force. The ``capture range'' refers to the range of
velocities which can be effectively cooled, and is defined by
$-v_E<v<v_E$, where the critical velocity $\pm v_E$ corresponds to the
maximum magnitude of the friction force.  In the weak probe regime and
in the limit of small probe detuning $\Delta_p\ll\Omega_c^2/\gamma_3$,
the critical velocity is approximately
$v_E\approx\Omega_c^2/(\gamma_3 k)$. 

\begin{figure}[H]
\centering
\includegraphics[scale=0.45]{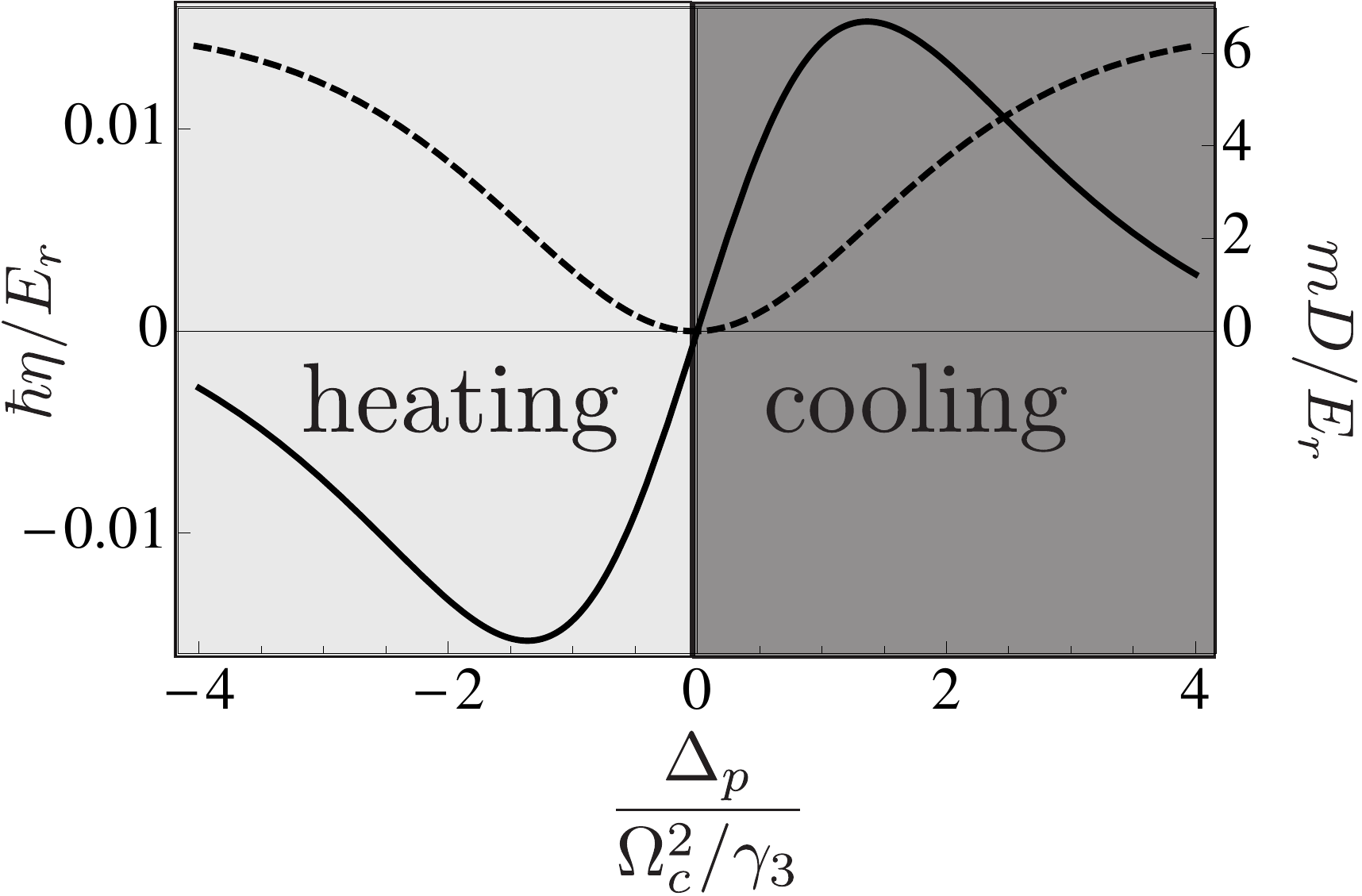}
\caption{Friction coefficient $\eta$ (solid line)
  and diffusion coefficient $D$ (dashed line) as a function of
  detuning. Both the friction and diffusion coefficients are zero at
  $\Delta_p=0$. The friction, $\eta$, reaches a maximum at
  approximately $\Delta_p=2\Omega_c^2/\gamma_3$, while the
  diffusion, $D$, is maximized at approximately $\Delta_p=\Omega_c$.
  The parameters used were, $\Omega_p=20E_r/\hbar$,
  $\Omega_c=400E_r/\hbar$ and $\gamma_3=2000E_r/\hbar$.}
\label{fig:rates}
\end{figure}

We find the friction coefficient
$\eta(x)$ averaged over one wavelength is given by
\begin{equation}
\begin{split}
  \eta&\equiv\overline{\eta(x)} \approx\frac{\hbar k^2 }{m}\frac{16
    \gamma _3\Delta _p \Omega _p^2 (\Omega_c^4-\Delta_p^4)} {\left(4
      (\Omega_c^2-\Delta_p^2)+\gamma _3^2 \Delta _p^2 \right){}^2}.
\end{split}
\label{eq:eta}
\end{equation}
The friction coefficient $\eta$ is plotted in Fig.~\ref{fig:rates}.
For blue detuning, $\Omega_c>\Delta_p>0$, the friction coefficient is
positive, $\eta>0$, which is consistent with the anticipated cooling
behavior discussed earlier; for red detuning $-\Omega_c<\Delta_p<0$,
we find a negative friction coefficient, $\eta<0$, and thus the atoms
are heated.

Now we turn to a discussion of the role of fluctuations which, in
concert with the friction forces just considered, determines the
steady-state temperature. There are
two main heating sources~\cite{PhysRevA.21.1606}: the first one
arises from the random momentum kicks that the atom receives when it
spontaneously emits a photon and recoils; the second arises from the
zero-point fluctuations of the atomic dipole moment (due to the
interaction with vacuum modes of the radiation field) in the presence
of electric field gradients~\cite{DeMarco}. This effect is also
intimately associated with spontaneous emission. In order to account
for both of these two processes, we include a stochastic term
$\hat\xi(t)$ in the equation of motion
\begin{equation}
\frac{d p}{dt}=-\eta p+\hat\xi(t)\,,
\label{eq:stoch}
\end{equation}
which satisfies the time average behavior of a white noise source,
{\it i.e.}  $\langle\hat\xi(t)\rangle=0$, and
$\langle\hat\xi(t)\hat\xi^\dagger(t')\rangle=2D\delta(t-t')$ where $D$ is the
diffusion constant~\cite{gardiner2004quantum}
\begin{equation}
2D=\frac{d\langle \hat p^2\rangle}{dt}-2\langle \hat p\rangle
  \langle\frac{d\hat p}{dt}\rangle\,. 
\end{equation}
$\mathcal N(u)$ depends on the specific quantum numbers of the
transition, and has a minor effect on the numerical value of 
the diffusion constant. Thus for simplicity, we will assume
the momentum kicks arising from spontaneous emission happen only along
the $\pm x$ directions and with equal probability, that is
$\mathcal{N}(u)=\frac{1}{2}\delta(u+1)+\frac{1}{2}\delta(u-1)$.  Using
this prescription, we obtain $D$ by solving the master equation
Eqn.(\ref{eq:EIT}) in the weak probe limit, giving
\begin{equation}
\begin{split}
  2D\approx\frac{4 (\hbar k)^2\gamma_3 \Omega_p^2 \Delta_p^2}
  {(\Omega_c^2-\Delta_p^2)^2+(\frac{\Delta_p\gamma_3}{2})^2}
\end{split}
\label{eq:diffusion}
\end{equation}

The equilibrium temperature is found by comparing the relative
magnitude of the cooling rate $\eta$ and the diffusion $D$:
\begin{eqnarray}
 k_B T&=&\frac{D}{m\eta}
  =\frac{\hbar}{2}\frac{(\gamma_3/2)^2+(\Omega_c^2/
  \Delta_p-\Delta_p)^2}{\Delta_p(\Omega_c^4/\Delta_p^4-1)}\nonumber\\
  &\approx&
  \frac{\hbar}{2}[\frac{1}{4}(\frac{\gamma_3}{\Omega_c^2})^2
\Delta_p^3+\Delta_p]
  \approx
  \frac{\hbar\Delta_p}{2}.
\label{eq:temperature}
\end{eqnarray}
We have validated this formula by establishing its consistency with full
semiclassical numerical simulations utilizing the
$c$-number Langevin method, as described in detail in \cite{supplementary}. For the numerical
simulations it was not necessary to make the weak probe
approximations, or to take the ultracold limit, as was required to
allow tractable analytic expressions to be derived in the formulas we
have just presented.

It should be emphasized that three conditions are required for
Eqn.~(\ref{eq:temperature}) to be valid. 
\begin{enumerate}
\item $\Delta_p\ll2\Omega_c^2/\gamma_3$. This ensures the detuning is
  confined to the blue side of the dark absorption dip.
\item $kv\ll \Delta_p$ for thermal velocities.  This ensures the
  Doppler shift is not too large in comparison with the probe
  detuning. This requirement corresponds to the constraint that the
  friction and diffusion coefficients should be constant and not
  depend on velocity.
\item $k_B T\gg E_r$, that is, the thermal energy exceeds the recoil
  energy. This is due to the fact that a semiclassical treatment has
  been used. Should this condition not be met, a fully quantum
  mechanical treatment of the motional wavefunction would be necessary.
\end{enumerate}
Although the first of these conditions is always required, it may be
possible to relax the second and third conditions and still achieve
laser cooling using this EIT approach. In the next section, we present
a numerical treatment which demonstrates this more general theory case.

%-------------------------------------------------------------------------------------------------------------------------------------------
\subsection{Quantum Mechanical Treatment}

In parameter regions where the predicted thermal energy from the
semiclassical treatment approaches the recoil energy, the
semiclassical method fails and the steady-state temperature must be
found by solving the density matrix quantum mechanically,
e.g. treating the external degrees of freedom (momentum and position)
also as operators.  However, without any simplification, it becomes
difficult to solve the dynamics due to the large number of relevant
basis states: if the momentum is in the range of $-N\hbar k$ to
$N\hbar k$ with a discrete step $\hbar k$, the density matrix scales
as $N^2$.  However, the Monte-Carlo wavefunction (MCWF)
method~\cite{Molmer:93}, which is based on the propagation of
stochastic differential equations, reduces the computation complexity
by requiring the storage of wavefunctions, which only scale as $N$, at
the cost of computing an ensemble of simulated trajectories. The MCWF
method has shown to be equivalent to the density matrix description
and has been applied to a wide variety of quantum optics problems,
including laser cooling.  Here, we use the MCWF method to accurately
determine the minimum reachable temperature.  Details of the numerical
procedure are presented in \cite{supplementary}.

\begin{figure}[H]
\centering
\includegraphics[scale=0.35]{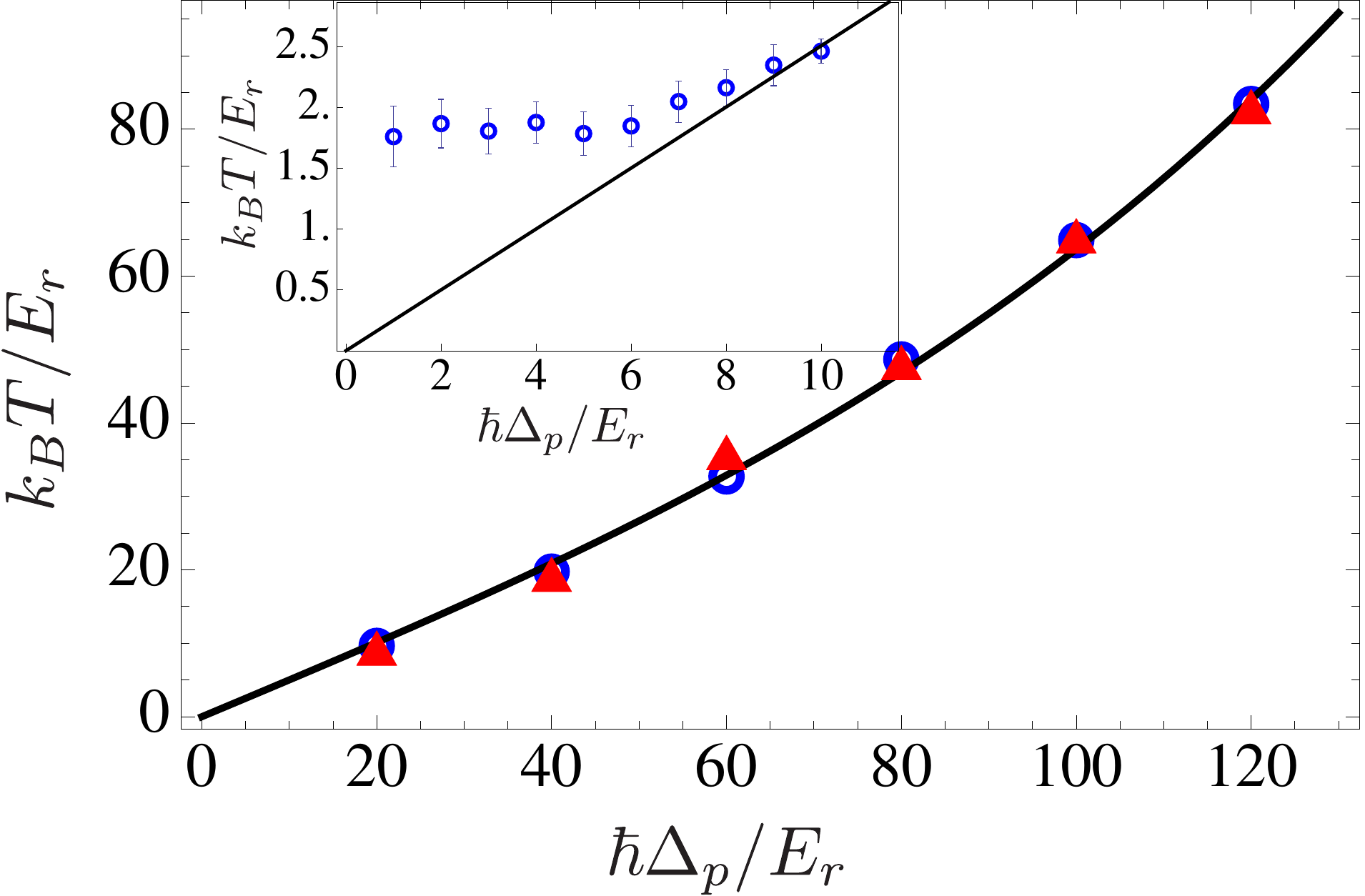}
\caption{ Final temperature reached as a function of the detuning. The
  parameters were $E_r/\hbar=5\rm{kHz}$, $\Omega_c=2\rm{MHz}=400 E_r/\hbar$,
  $\Omega_p=100\rm{kHz}=20 E_r/\hbar$, $\gamma_3=10\rm{MHz}=2000E_r/\hbar$.  The m
  truncated momentum basis ranged from $-50\hbar k$ to $50\hbar k$.
  The analytical
  formula (solid line), semiclassical numerical results (triangle) and
  MCWF numerical results (circle) are compared.  They agree well when
  the detuning is significantly larger than the recoil energy, but
  discrepancies appear when the detuning is near the recoil energy, as
  shown in the inset plot.
  The standard errors for the small detuning points are in the order of 0.1 as shown by the error bars in the inset, 
  while the standard errors for large detunings points are in the order of 0.01, 
  which are too small to be depicted in the large plot.}
\label{fig:temperature}
\end{figure}

In order to determine the minimum reachable temperature and validate
the consistency of the various approximations made in deriving the
analytic formulas, we performed a numerical study using a set of
experimentally reasonable parameters.  Fig.~\ref{fig:temperature}
summarizes the main results.  It shows comparisons between the
analytical model and two different numerical solutions, one based on
the $c$-number Langevin treatment of the semiclassical equations, and
the other based on the MCWF treatment of the full quantum
solutions. All of them are consistent for parameters in which the
equilibrium temperature is much greater than the recoil temperature.
Discrepancies between the analytical formula and the result from MCWF
are apparent when $\Delta_p\sim E_r$, as anticipated in the discussion
presented in Sec.~\ref{sec:mechanism}. We note that when $\Delta_p$
approaches zero the observed temperature is close to the recoil limit
rather than below it. This is in direct contradiction to
Eqn.~(\ref{eq:temperature}) that predicts no lower limit.

\section{Atomic Motion}
\label{sec:trapping}

In the weak probe regime, the conservative force that arises from the
standing wave light field does not play a significant role, but this
ceases to be true when strong probe fields are considered. In general,
the induced lattice potential must be included.  The important
question is then: can an atom can be trapped in and confined to
a single site of the lattice, and if so, how long will it remain
trapped when subjected to the momentum kicks induced by random
fluctuations? The random fluctuations compete with friction forces
that cool the atom and may result in localization for considerable
periods of time. One may expect that the effectiveness of a trap in
confining an atom would be determined by the ratio of the lattice
depth to the temperature at equilibrium. Assuming a small Doppler shift,
$kv\ll\Delta_p$, the periodic potential $V(x)$ is given by 
\begin{equation}
 V(x) =\frac{16\hbar \Delta _p \tilde\Omega _p^2 
\left(\Omega _c^2-\Delta _p^2+4 \tilde\Omega _p^2\right)}
 {4 \left(2 \Delta _p^2 \left(4 \tilde\Omega _p^2-\Omega _c^2
\right)+\left(\Omega _c^2+4 \tilde\Omega _p^2\right){}^2+\Delta _p^4
\right)+\gamma _3^2 \Delta _p^2}
\label{eq:trapping1}
\end{equation}
where $\tilde\Omega_p=\Omega_p\cos(kx)$.
This expression can be simplified in the regime of the usual operating
conditions of small probe detuning, $\Delta_p\ll\Omega_c^2/\gamma_3$,
as
\begin{equation}
 V(x) \approx\frac{\hbar\Delta_p}{1+(\Omega_c/2\tilde\Omega_p)^2}
\label{eq:trapping2}
\end{equation}
The lattice depth $V_d$ can be calculated by $V_d=|V(0)-V(\pi/2k)|$.
The friction coefficient and the temperature in this strong probe
limit are difficult to derive analytically therefore we calculate them as
observables from the numerical simulations.

\begin{figure}[H]
\centering
\includegraphics[scale=0.45]{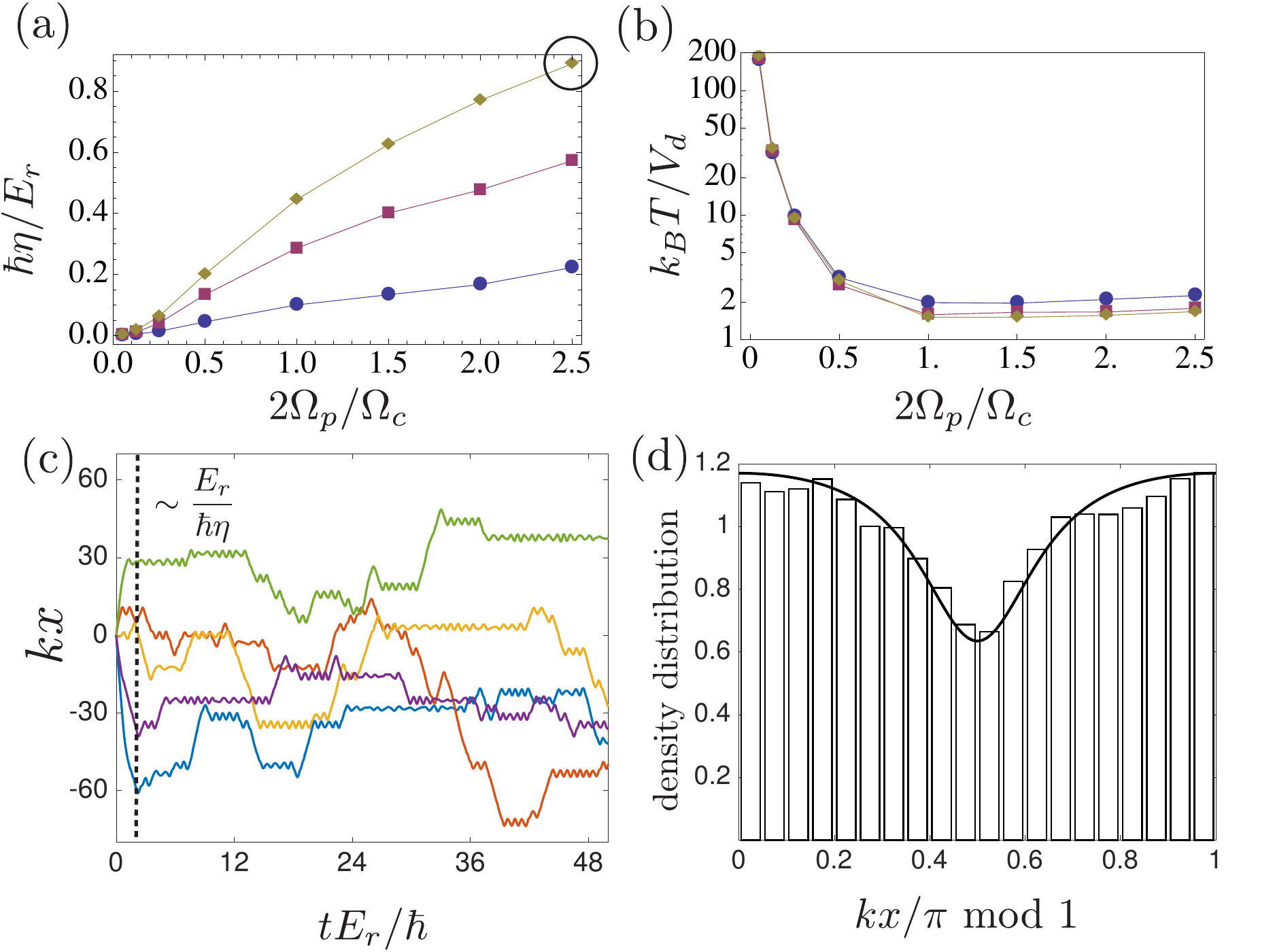}
\caption{(a)The friction coefficient as a function
  of $2\Omega_p/\Omega_c$, with the parameters $\gamma_3=2000E_r/\hbar$ and
  $\Omega_c=400E_r/\hbar$ at three detunings within the absorption dip; they
  are $\Delta_p=10E_r/\hbar$ (round), $\Delta_p=30E_r/\hbar$ (square), and
  $\Delta_p=50E_r/\hbar$ (diamond) respectively.  (b)The ratio between the
  lattice depth and the final kinetic energy as a function of
  $2\Omega_p/\Omega_c$, with the same simulation parameters as (a).
  (c) This subfigure shows $5$ trajectories, representing the atom
  position $x$ as a function of time.  The simulation parameters are
  extracted from the circle in \text{a}.  At about $t\sim1/\eta$, 
  as shown by the dashed line, the atom is cooled down.  
  After short periods of weak trapping within
  the wells of the lattice, the atom may gain sufficient energy to
  propagate over multiple sites before being cooled and recaptured.
 (d) This subfigure shows the position dependence of the density distribution, 
 with the same parameters as used in (c).
 The histogram is obtained from the simulations 
 by recording the position of each atom after reaching equilibrium.
 The solid line represents the density distribution that is analytically derived by 
 assuming that the thermodynamics of the equilibrated atomic gas 
 obeys the Boltzmann distribution.
}
\label{fig:trapping}
\end{figure}

Fig.~\ref{fig:trapping}(a) and Fig.~\ref{fig:trapping}(b) show $\eta$
and $k_B T/V_d$ as a function of the probe intensity $2\Omega_p$
ranging from the weak probe limit $\Omega_c/20$ to the strong probe
limit $2.5\Omega_c$.  While $\eta$ grows quadratically with $\Omega_p$
in the weak probe regime, it grows at a slower rate when $\Omega_p$
becomes comparable or larger than $\Omega_c$.  In the weak probe
limit, the final kinetic energy is not affected by $\Omega_p$ and the
optical lattice potential effectively vanishes; in the strong probe
limit, the lattice depth becomes large but the final kinetic energy
also increases dramatically.  The comparison between the equilibrium
temperature and the lattice depth is found to be $k_BT>V_d$ for the
conditions under consideration.  The trajectories, obtained from the
simulations, represent the atoms position $x$ as a function of time
$t$.  

Each single atom starts with a velocity within the capture rage
selected randomly from a high-temperature Maxwellian
distribution. While its position is observed to change significantly
at the beginning, gradually, the atom begins to be cooled and the
position is observed to change at a slower rate. Finally, at about
$t\sim1/\eta$, the atom is cooled to steady-state and becomes weakly
trapped in the induced lattice. These general features are illustrated
in Fig.~\ref{fig:trapping}(c) where trajectories are clearly visible
that oscillate back and forth, demonstrating the trapping of atoms in
a single site. However, since the temperature is never significantly
below the lattice depth, this trapping is transient, and the diffusion
drives the atom out of a site, after which it may gain
sufficient energy to fly over several sites before being recaptured.
Fig.~\ref{fig:trapping}(d) shows the position-dependent density distribution.
The histogram is made by recording the position of each atom after reaching equilibrium in the simulation.
The solid line represents the analytically derived density distribution.
By assuming that the thermodynamics of the equilibrated atomic gas 
obeys the Boltzmann distribution, 
the spatially variant density distribution is found to be proportional to $\exp(-V(x)/k_B T)$,
where $k_B T\approx 70E_r=1.75V_d$ is read from Fig.~\ref{fig:trapping}(b).
These density distributions found by the two methods agree well.
Fig.~\ref{fig:trapping}(d) apparently shows that the atoms are not completely trapped at the bottom of the lattice, 
which is consistent with the conclusions drawn from Fig.~\ref{fig:trapping}(b) and Fig.~\ref{fig:trapping}(c) and 
suggests that the trapping effect by this mechanism is weak.

%-------------------------------------------------------------------------------------------------------------------------------------------
%-------------------------------------------------------------------------------------------------------------------------------------------
\section{Conclusions}
\label{sec:conclusion}

We have proposed and analyzed a novel cooling mechanism based on a
Lambda-type EIT system, where heating of the atoms is largely
suppressed by destructive quantum interference.  We have showed
in the weak probe beam regime that, by blue detuning the laser beams it is
possible to take advantage of the EIT reduced absorption window to
cool down the atoms, and to reach sub-Doppler final temperatures set
by the detuning instead of the atoms' decay rate. We developed and
analyzed the results of a model obtained using a semiclassical
approach, which we compared with the solutions of the full quantum
dynamics using a MCWF method.  Discrepancies appeared when the
detuning approached zero, where the final temperature was comparable
to the recoil energy.  In the strong probe beam region, we derived the
depth of the optical lattice potential experienced by the atom, and
compared it with the final temperatures.  By computing the ratio of
the final temperature to the lattice depth, we found that the atoms
can be weakly trapped, which is
consistent with the observed behavior of the atomic motion in the
simulations.

In our discussion, we assumed that the atomic gas was dilute.
In fact, it is often the case in laser-cooling experiments that the atomic gas is not in this regime, but is optically thick.
Therefore, laser beams may be drastically attenuated as they propagate,
and may be subject to strong non-linear optical effects.
Consequently, in real systems, fewer atoms may be subject to the intended dissipative forces,
which may significantly modify the efficacy and achievable temperatures.
This adverse effect arising from photon absorption and dispersion in the optically thick sample 
could be significantly mitigated by the use of the absorption dip in
the EIT transparency window as we have proposed. Furthermore, the degree of
transparency is directly controllable by tuning the probe laser. 
It is worthy to note that although the cooling rate per atom is reduced 
as a result of the decrease of the detuning, 
it is possible for
the total cooling rate for the whole system to remain roughly unchanged 
since more atoms may be cooled in such a transparent system.

Moreover, multi-atom interactions can be important in a dense gas, 
including both ground state collisions, dipole-dipole interactions, 
light-assisted collisions, as well as the macroscopic radiation pressure 
forces that can limit the total number of atoms that can be trapped and cooled. 
Many of these interactions are mediated by photon emission, 
and hence could be largely suppressed by using the EIT absorption dip~\cite{chang2014quantum, gibble1992improved}. 
Although we have not included
calculations of multi-atom effects in this paper, the topic will be an
interesting extension for future work.
%In our discussion, we assumed that the atomic gas was so dilute that
%the atoms in the sample could be considered to be completely
%independent.  However, in many conventional laser cooling systems,
%atom- effects are important and often significantly modify the
%efficacy and achievable temperatures as seen in experiments. In the
%case of an optically thick quantum gas, laser beams are drastically
%attenuated as they propagate, leading to spatial dependence of the
%Rabi frequencies, and consequently additional forces and
%inefficiencies in the laser cooling dynamics.
%Moreover, atom-atom interactions can be important in dense gas, 
%including both ground state collisions,
%dipole-dipole interactions, light-assisted collisions, as well as the
%macroscopic radiation pressure forces that can limit the total number
%of atoms that can be trapped and cooled. Many of these adverse effects
%are due to the photon absorption in a dense sample, and consequently
%could be significantly mitigated by the use of the absorption dip in
%the EIT window as we have proposed. Furthermore, the degree of
%transparency is directly controllable by tuning the probe laser at the
%cost of altering the cooling rate. Although we have not included
%calculations of multi-atom effects in this paper, the topic will be an
%interesting extension for future work.

In this paper, we have limited the discussion to one-dimensional (1D)  systems, even
though typical laser cooling applications often require 2D or 3D
cooling. It is likely that there are extensions of the model we have
presented to laser cooling in higher dimensions. Furthermore, even
though the capture range is limited by the Rabi frequency of the
coupling laser and the linewidth of the excited state, strategies
could be implemented in which these are swept in time to capture large
numbers of hot atoms, and later to cool them to very low
temperatures. It will be interesting to explore such general
approaches.

We acknowledge helpful discussions with Minghui Xu.
The material is based upon work supported by
the National Science Foundation under Grant Numbers PHY1125844 \& PHY1404263, 
the National Institute of Standards and Technology, 
the DARPA QuASAR program, and the Air Force Office of Scientific Research under Grant Number FA9550-14-1-0327.

%\bibliography{Papbib}


\widetext

\appendix
\section{$C$-Number Langevin Method }
\label{App:AppendixA}
In this method, the atomic momentum and position are treated
semiclassically, and the dynamical equation of motion for the atomic
momentum is augmented with a stochastic term. The weight of the
stochastic fluctuation is set so that the resulting momentum diffusion
agrees with that arising from quantum noise in a full quantum
treatment of the same problem. In practice, this means that a single
atom evolves according to Eqn.~(\ref{eq:Langevin}), in which the
differential equation for momentum $p$ contains a Gaussian fluctuating
stochastic process $\sqrt{2D}\,\xi^{\rm class}$ that replaces the
quantum noise operator $\hat\xi$ in the quantum Langevin equation (
originally given in Eqn.~(7) in the paper). The closed set of Ito
stochastic differential equations to solve numerically are then given
by
\begin{eqnarray}
\frac{dp}{dt}&=&4 \sin(kx)\Omega_p(\rho_{13}-\rho_{31})
+\sqrt{2D}\,\xi^{\text{class}}\nonumber\\
\frac{dx}{dt}&=& \frac{p}{m}\nonumber\\
\frac{d\rho_{23}}{dt}&=&-\frac{\gamma_3}{2}\rho_{23}
+i(\rho_{33}-\rho_{22})\Omega_c-2 i \rho_{21}\Omega_p\cos(kx)\nonumber\\
\frac{d\rho_{12}}{dt}&=&i\Delta_p\rho_{12}-i\rho_{13}\Omega_c
+2i\rho_{32}\Omega_p\cos(kx)\nonumber\\
\frac{d\rho_{13}}{dt}&=&i\Delta_p\rho_{13}-\frac{\gamma_3}{2}
\rho_{13}-i\rho_{12}\Omega_c+2i(\rho_{33}-\rho_{11})\Omega_p\cos(kx)
\nonumber\\
\frac{d\rho_{22}}{dt}&=&-i\Omega_c(\rho_{23}-\rho_{32})\nonumber\\
\frac{d\rho_{33}}{dt}&=&-\gamma_3\rho_{33}+i\Omega_c(\rho_{23}
-\rho_{32})+2i(\rho_{13}-\rho_{31})\Omega_p\cos(kx)
\label{eq:Langevin}
\end{eqnarray}

In order to solve these equations, a number of alternative algorithms
are available, such as the Milstein method and the Runge-Kutta
method. We have implemented various numerical algorithms and do not
find our results to depend on the choice. In this paper, we have
presented numerical calculations using the Euler-Maruyama method,
which is a simple generalization of the Euler method for numerical
integration to treat stochastic differential equations. Explicitly,
the differential equation for the momentum is treated in the following
way. Advancing the momentum over a numerical integration step of size
$\Delta t$, we evaluate
\begin{equation}
\Delta p=4 \sin(kx)\Omega_p(\rho_{13}-\rho_{31})\Delta
t+\sqrt{2D}\,\Delta\xi^{\text{class}}\,.
\end{equation}
The random variable $\Delta\xi^{\text{noise}}$ for each time step is
treated as statistically independent and is found by sampling (using a
numerically produced quasi-random number) from a normal distribution
with zero mean and variance $\Delta t$.  Together with the properties
of the density matrix $\Tr\rho=1$ and $\rho^\dag=\rho$, the time
dependence of all elements of the density matrix can be tracked.

\section{Monte-Carlo Wave Function (MCWF) Method} 
\label{App:AppendixB}
The numerical procedure used to solve the quantum dynamics using the
MCWF method is the following: the time-dependent wave function is
expanded in terms of the basis $|i,\, p\rangle$. Here $i$ denotes the
internal states and $p$ accounts for the quantized momentum along $x$
axis.  The momentum grid is specified as a family of momenta each
separated by~$\hbar k$, and with a central offset $\tilde p$, giving a
grid of momentum basis states that range from $-N\hbar k+\tilde{p}$ to
$N\hbar k+\tilde{p}$. We expand a general quantum state
$|\psi(t)\bigl>$ in this basis as
\begin{equation}
|\psi(t)\rangle=\sum_{ \, p_n=-N\hbar k}^{N \hbar k} \sum_{i=1}^{3}c_{i,n}(t) 
|i, \, \tilde{p}+p_n\rangle
\end{equation}
To account for the dissipative processes induced by spontaneous
emission, the non-Hermitian evolution is added to the Hamiltonian to
give
\begin{eqnarray}
  H_{\rm eff}&=&\sum_{p_n=-N\hbar k}^{N \hbar k}\Big\{\sum_{i=1}^{3}
  \frac{(\tilde{p}+p_n)^2}{2m}|i, \, \tilde{p}+p_n\rangle\langle i,
  \, \tilde{p}+p_n|+\hbar\Delta_p|1, \tilde{p}+p_n\rangle\langle 1,\tilde{p}+p_n|\nonumber\\
  &&+\hbar\Omega_p\big(
  |1, \tilde{p}+p_n\rangle\langle3,
  \tilde{p}+p_{n+1}|+|1, \tilde{p}+p_n\rangle\langle3,
  \tilde{p}+p_{n-1}|+{\rm h.c.}
  \big)\nonumber\\
  &&+\hbar\Omega_c\big(
  |2, \tilde{p}+p_n\rangle\langle3,\tilde{p}+p_{n}|+{\rm h.c.}
  \big)
  -\frac{i\hbar\gamma_3}{2}|3, \tilde{p}+p_n\rangle\langle3, \tilde{p}+p_n|\Big\}
\end{eqnarray}
where ${\rm h.c.}$ denotes the hermitian conjugate.

The wave function is evolved under Schrodinger's equation, giving
coupled differential equations for the coefficients $c_{i,n}(t)$
\begin{eqnarray}
\frac{d\, c_{1,n}(t)}{dt}&=&-i\Big(\frac{(\tilde{p}+n
\hbar k)^2}{2m\hbar}+\Delta_p\Big)c_{1,n}(t)
-i\Omega_p\big(c_{3,n+1}(t)+c_{3,n-1}(t)\big) \nonumber\\
\frac{d\, c_{2,n}(t)}{dt}&=&-i\frac{(\tilde{p}+n\hbar k)^2}
{2m\hbar}c_{2,n}(t)-i\Omega_c c_{3,n}(t) \nonumber\\
\frac{d\, c_{3,n}(t)}{dt}&=&-i\Big(\frac{(\tilde{p}+n\hbar k)^2}
{2m\hbar}-\frac{i\gamma_3}{2}\Big)c_{3,n}(t)
-i\Omega_p\big(c_{1,n+1}(t)+c_{1,n-1}(t)\big)-i\Omega_c c_{2,n}(t)
\label{eq:jump}
\end{eqnarray}
Numerically, we apply a second-order Runge-Kutta method to solve
Eqn.~(\ref{eq:jump}). A quantum jump (a single dissipative event)
occurs at a time $t$ when the norm of the wave function becomes
smaller than a fixed initially produced random number that is drawn
from a uniform distribution in the interval $[0,\,1]$.  The action of
the quantum jump on the wavefunction is given by applying the rules
\begin{eqnarray}
c_{1,n}(t)&&\rightarrow c_{1,n}(t)=c_{3,n}(t)\nonumber\\
c_{2,n}(t)&&\rightarrow c_{2,n}(t)=0\nonumber\\
c_{3,n}(t)&&\rightarrow c_{3,n}(t)=0\nonumber\\
\tilde p(t)&&\rightarrow \tilde p(t)=\tilde p(t)\pm \hbar k
\end{eqnarray}
and the wave function is subsequently renormalized to unit
norm. Whether the $+$ or $-$ is used is determined by flipping a coin,
coinciding with the random direction of the emitted photon and leading
to momentum diffusion in the quantum treatment. After each emission
event, the atom is found in its ground state and the momentum
distribution has been shifted by an amount $\hbar k$.

\end{document}